\documentclass[12pt,preprint]{aastex}

\shorttitle{CO Oscillator Strengths}
\shortauthors{Eidelsberg et al.}

\begin{document}
\title{Oscillator Strengths and Predissociation Rates 
for Rydberg Transitions in $^{12}$C$^{16}$O, 
$^{13}$C$^{16}$O, and $^{13}$C$^{18}$O Involving the $E~^1\Pi$, 
$B~^1\Sigma^+$, and $W~^1\Pi$ States}
\author{M. Eidelsberg\altaffilmark{1}, Y. Sheffer\altaffilmark{2},
S.R. Federman\altaffilmark{2}, 
J.L. Lemaire\altaffilmark{1}$^,$\altaffilmark{3}, 
J.H. Fillion\altaffilmark{1}$^,$\altaffilmark{3}, 
F. Rostas\altaffilmark{1}, and J. Ruiz\altaffilmark{4}}
\altaffiltext{1}{Observatoire de Paris-Meudon, LERMA UMR8112 du CNRS, 
France; michele.eidelsberg@obspm.fr; jean-louis.lemaire@obspm.fr; 
Jean-Hugues.fillion@lamap.u-cergy.fr; francois.rostas@obspm.fr.}
\altaffiltext{2}{Department of Physics and Astronomy, University of Toledo,
Toledo, OH 43606; \linebreak
ysheffer@physics.utoledo.edu; steven.federman@utoledo.edu.}
\altaffiltext{3}{Universit\'e de Cergy-Pontoise, LERMA UMR8112 du 
CNRS, France.}
\altaffiltext{4}{Department of Applied Physics I, Universidad de 
M\'{a}laga, 29071-M\'{a}laga, Spain; jruiz@uma.es.}

\begin{abstract}
One of the processes controlling the interstellar 
CO abundance and the ratio of its isotopologues is photodissociation.  
Accurate oscillator strengths and predissociation rates 
for Rydberg transitions are needed for modeling this process.  
We present results on absorption from the $E~^1\Pi-X~^1\Sigma^+$ (1-0) 
and $B~^1\Sigma^+-X~^1\Sigma^+$ (6-0) bands at 1051 and 1002 \AA, 
respectively, and the vibrational progression $W~^1\Pi-X~^1\Sigma^+$ 
($v^{\prime}$-0) bands with $v^{\prime}$ $=$ 0 to 3 at 972, 956, 
941, and 925 \AA, respectively.  The corresponding spectra were 
acquired at the high resolution (R $\approx$ 30,000) SU5
beam line at the Super ACO Synchrotron in Orsay, France.  Spectra
were obtained for the $^{12}$C$^{16}$O, $^{13}$C$^{16}$O, and
$^{13}$C$^{18}$O isotopologues.  These represent the most complete 
set of measurements available.  Comparison is made with earlier 
results, both empirical and theoretical.  While earlier 
determinations of oscillator strengths based on absorption 
from synchrotron radiation tend to be somewhat smaller than ours, 
the suite of measurements from a variety of techniques agree 
for the most part considering the mutual 
uncertainties.  For the bands studied here, their relative 
weakness, or their significant line widths arising from 
predissociation, minimizes potential problems from large optical 
depths at line center in absorption measurements.  Predissociating 
line widths could generally be extracted from the spectra thanks 
to the profile simulations used in the analysis.  In many cases, 
these simulations allowed us to consider $e$ and $f$ parity 
levels separately and to determine the dependence of the 
width on rotational quantum number, $J$.  Our results 
are consistent with earlier determinations, 
especially the widths inferred from laser experiments.
\end{abstract}


\section{Introduction}

The photochemistry of CO plays an important role in photon 
dominated regions (PDRs) of interstellar clouds, circumstellar 
shells surrounding AGB stars, planetary nebulae, circumstellar 
disks around young stars, and comets.  Models of these regions 
(e.g., van Dishoeck \& Black 1988; Sternberg \& Dalgarno 1995; 
Hollenbach \& Tielens 1997, and references therein) use 
observational data on CO transitions to extract physical 
conditions.  The models rely on oscillator strengths for 
dissociating transitions (e.g., Letzelter et al. 1987; 
Eidelsberg et al. 1991; Chan, Cooper, \& Brion 1993; Federman et 
al. 2001; Sheffer, Federman, \& Andersson 
2003; Eidelsberg et al. 2004a).  While a consensus is emerging 
for oscillator strengths pertaining to many of the important 
electronic transitions (Federman et al. 2001; Eidelsberg et al. 
2004a), the large number of transitions leading to photodissociation 
warrants a careful assessment of as many 
measurements as possible.  Here we 
present results on the $E~^1\Pi-X~^1\Sigma^+$ (1-0), 
$B~^1\Sigma^+-X~^1\Sigma^+$ (6-0), and core excited 
$W$($A^2\Pi^+$)$~^1\Pi-X~^1\Sigma^+$ ($v^{\prime}$-0) bands with 
$v^{\prime}$ $=$ 0 to 3 for $^{12}$C$^{16}$O, $^{13}$C$^{16}$O, 
and $^{13}$C$^{18}$O acquired at the Super ACO synchrotron source 
in Orsay, France.

These results are a continuation of our earlier efforts.  
Federman et al. (2001) studied Rydberg transitions from the 
ground vibrational level in $^{12}$C$^{16}$O between 1150 
\AA\ and about 1075 \AA.  The limitation on the short 
wavelength end was the presence of LiF windows for the gas 
cell.  More recently, a system without windows based on 
differential pumping allowed us to measure CO absorption from 
bands with wavelengths approaching 912 \AA, the Lyman limit.  
Eidelsberg et al. (2004a) analyzed the strongly interacting 
bands [$K-X$ (0-0), $L^{\prime}-X$ (1-0), and 
$L-X$ (0-0)] near 970 \AA\ in $^{12}$C$^{16}$O, 
$^{13}$C$^{16}$O, and $^{13}$C$^{18}$O.  We now turn our 
attention to the other transitions involving Rydberg states 
that were acquired during the same synchrotron runs.  
The transitions discussed in the present paper are seen in 
interstellar spectra (Sheffer et al. 2003) and are important 
for CO photodissociation (van Dishoeck \& Black 1988).

\section{Experimental Details and Data Analysis}

Since our spectra were acquired during the runs described by 
Eidelsberg et al. (2004a), we only provide a brief overview 
here.  The measurements were obtained on the Super ACO 
synchrotron ring using the high spectral resolution SU5 
beam line (Nahon et al. 2001a, b).  The spectrometer 
consisted of a 2400 lines mm$^{-1}$ SiC grating blazed at 
13 eV.  The entrance and exit slits were set at 35 or 
25 $\mu$m, yielding respective instrumental widths of 
30 or 20 m\AA, as deduced from the analysis based 
on profile simulation described below.  
The vacuum ultraviolet (VUV) light emerged 
from the spectrometer and passed through a windowless gas 
cell; differential pumping maintained the ultra-high 
vacuum in the spectrometer and storage ring.  High purity 
CO gas ($^{12}$C$^{16}$O $-$ Alphagaz, 99.997\%; 
$^{13}$C$^{16}$O $-$ Eurisotop, 99.1\% $^{13}$C, 99.95\% 
$^{16}$O; $^{13}$C$^{18}$O $-$ Isotech, 98.8\% $^{13}$C, 
94.9\% $^{18}$O) continuously flowed through the cell.  
The pressure in the cell, typically 1 to 10 mTorr except 
for the weakest bands, was measured with a capacitance 
gauge whose full scale was 10 Torr.  The stability of the 
pressure was checked by monitoring a cold cathode 
ionization gauge in the differentially pumped 
section where typical pressures were 10$^{-6}$ Torr.  
The VUV flux exiting the gas cell was detected with 
a photomultiplier tube (EMI 9558QB) placed after a 
sodium salicylate coated window acting as a scintillator.  
The digitized signal from the PMT was recorded on a PC.

Oscillator strengths ($f$-values) were derived through 
syntheses of the measured absorption bands.  
As noted in Eidelsberg et al. (2004a), two independent codes 
were used to check for consistency, where agreement at the 
2\% level was achieved.  Previous analyses of absorption 
from Rydberg bands of CO with the Toledo-based code 
included the experimental work of Federman et al. (2001) 
and the interstellar study of Sheffer et al. (2003).  The 
code developed in Meudon was used on $A-X$ bands 
(Eidelsberg et al. 1999) and intersystem transitions 
(Rostas et al. 2000).  The synthetic spectra were based on 
tabulated spectroscopic data (Eidelsberg et al. 1991).   
Each synthetic spectrum was adjusted 
to match the experimental one in a non-linear least-squares 
fitting procedure with band oscillator strength, line width 
(instrumental, thermal, and predissociation), and wavelength 
offset as free parameters.  Initial fits revealed that the 
thermal width was consistent with Doppler broadening at 
295 K and was held constant in subsequent syntheses.  
Predissociation widths of 0.34 m\AA\ for the $E-X$ (1-0) 
band in $^{12}$C$^{16}$O, 0.27 m\AA\ in $^{13}$C$^{16}$O, 
and 0.18 m\AA\ for $^{13}$C$^{18}$O 
were taken from Ubachs, Velchev, and Cacciani (2000).  
(Hereafter we use the shorthand $E$1 because absorption 
examined in the present paper always involves the ground 
vibrational level.)  For the $W$0, $W$2, and 
$W$3 bands, a $J$-dependent width had to be introduced in 
the synthetic spectra in order to reproduce the observations, 
in keeping with the results of Drabbels et al. (1993) and 
Eikema, Hogervorst, and Ubachs (1994).  

The situation for the $W$2 bands required special care.  
According to Eikema et al. (1994), the 
predissociation width is independent of $J$ until $J$ $=$ 
6, but dependent on $J$ for $J$ $>$ 6 for $e$ and $f$ 
components in all isotopologues.  We first determined the 
$J$-independent part of the width from lines up to Q(6) and 
P(3), and then fitted the whole band to derive the 
$J$-dependent part and the $f$-value.

The first step in the analysis was to obtain an accurate 
measure of the CO column in the differentially-pumped 
system.  This was achieved through profile fits of the 
$E$0 band near 1075 \AA\ for each isotopologue.  Since the 
3$p\pi$ ($v^{\prime}$ $=$ 0) state is relatively isolated 
and can be considered essentially free of perturbations, the 
band $f$-value was assumed to be the same for the three 
isotopologues.  Predissociation widths of 0.067 m\AA\ for 
$^{12}$C$^{16}$O, 0.049 m\AA\ for $^{13}$C$^{16}$O, and 
0.047 m\AA\ for $^{13}$C$^{18}$O (Cacciani et al. 1998) were 
used as input.  The pressure deduced from the fit to the 
$E$0 band was typically within 10\% of the value 
given by the pressure gauge.  
The deduced pressure was then adopted in syntheses of the bands 
of primary interest scanned with the same set of conditions.  
For the high pressures needed to measure absorption from the 
$E$1 band of $^{13}$C$^{18}$O -- 24 to 58 mTorr, the values 
from the pressure gauge were used.

Examples of our fits for $E$1, $W0$ and $W$3, and $B$6 bands 
appear in Figs. 1 to 3, respectively.  The ordering is based on
predissociation width, from narrowest to broadest.  
These figures reveal that several of the bands experience 
perturbations.  In Fig. 1 accidental predissociation at 
$J_e$ $=$ 7 for $^{13}$C$^{16}$O (and $^{12}$C$^{16}$O) 
(Stark et al. 1992; Cacciani, Hogerworst, \& Ubachs 1995) gives
rise to extra lines appearing with R(6) and P(8).  
In $^{13}$C$^{18}$O it appears at $J_e$ $=$ 1 and 5 
(Ubachs et al. 2000) and gives 
rise to line shifts and broadening that are barely 
visible in our spectra.  These interactions arise from 
spin-orbit coupling between $E~^1\Pi$ ($v$ $=$ 1) and 
$k~^3\Pi$ ($v$ $=$ 6).  As for Fig. 3, the absorption 
bands at 1002.56 \AA\ in $^{12}$C$^{16}$O, at 
1002.81 \AA\ in $^{13}$C$^{16}$O, and at 1003.21 \AA\ in 
$^{13}$C$^{18}$O have been attributed to a transition between 
the ground state $X~^1\Sigma^+$ $v^{\prime\prime}$ $=$ 0 and 
a resonant state due to the strong coupling between the 
3$s\sigma B~^1\Sigma^+$ Rydberg state and the repulsive 
$D^{\prime}~^1\Sigma^+$ valence state (modeled by Tchang-Brillet 
et al. 1992; Monnerville \& Robbe 1994; Li et al. 1997; Andric et 
al. 2004).  This resonant state has been discussed in detail and 
identified by its Rydberg character to the vibrational level 
$v^{\prime}$ $=$ 6 of the $B~^1\Sigma^+$ state by Eidelsberg 
et al. (2004b).  As predicted by Andric et al. (2004), the 
broadening of these bands varies widely from one isotopologue to 
another.  Very weak and diffuse absorption 
bands involving the $v^{\prime}$ $=$ 4 and 5 levels 
have been observed recently (Baker 2005) below the 
avoided crossing of the diabatic potential curves.  $B~^1\Sigma^+$ 
$v^{\prime}$ $=$ 6 is thus the first resonance state to be observed 
above the avoided crossing.  Two other processes 
affecting the $f$-values for $B$6 and $W$1 in 
$^{13}$C$^{18}$O are discussed in the next section.

\section{Results and Discussion}

Our results are presented in Table 1 ($f$-values) and 
Table 2 (predissociation rates, $k_p$) along with values from 
earlier studies.  The (FWHM) Lorentzian line widths ($\Gamma_L$) 
that were determined from the spectra are linked to the 
corresponding upper level lifetimes ($\tau$) and transition 
rates ($k$) by the well known relation, 
$\Gamma_L$ = 1/(2$\pi \tau$) = (1/2$\pi$)$k$ ($k$ $=$ 
$\Gamma_L$(m\AA)~[1000/$\lambda$(\AA)]$^2$~$1.885\times10^{10}$ 
s$^{-1}$, where $\lambda$ is the band origin for the transition).  
The transition rate $k$ is the sum of the radiative and 
predissociation parts ($k_r$ and $k_p$, respectively).  
In the present experiments, the fact that $\Gamma_L$ can be 
measured at all implies that $k_p$ is considerably larger 
than $k_r$.  

Previous empirical $f$-values are based 
on synchrotron measurements (Eidelsberg et al. 1991; 
Stark et al. 1991, 1992, 1993; Yoshino et al. 1995), 
electron-energy-loss spectroscopy (Chan et al. 
1993; Zhong et al. 1997), and analysis of 
interstellar spectra (Sheffer et al. 2003).  As for the 
predissociation rates, an order of magnitude was first obtained 
by Eidelsberg et al. (1991) from widths evaluated from absorption 
features recorded in low-resolution synchrotron spectra.  Subsequent 
efforts (Levelt, Ubachs, \& Hogervorst 1992a,b; Drabbels et al. 1993; 
Eikema et al. 1994; Komatsu et al. 1995) were based on 
high-resolution line width measurements using laser techniques.  

Table 1 lists the results on oscillator strength.  
Columns 1-3 give the transition, wavelength of the band origin, 
and isotopologue, respectively.  Our $f$-values 
appear in the fourth column, and those from previous 
studies are shown in the remaining ones.  We 
indicate temperature for the measurements.  
For the transitions examined here, there generally is 
no difference in $f$-values among isotopologues within 
the uncertainty of our measurements.  

One exception involves the $B$6 band of $^{13}$C$^{18}$O, 
the $f$-value of which is significantly smaller than that 
of the other isotopologues.  It is to be noted that 
the 3$d\sigma-X~^1\Sigma^+$ (0-0) band appears 
on the blue wing of the $B$6 band in $^{13}$C$^{18}$O, 
while it is not seen in the other two isotopologues at the 
pressures used here ($<$ 24 mTorr).  As was shown in the 
detailed study of Eidelsberg et al. (2004b), the 3$d\sigma$0 
band borrows its intensity from the $B$6 band.  This effect is 
weak for the first two isotopologues, allowing this band to be 
seen only at high column density.  For the $B$6 band of 
$^{13}$C$^{18}$O, it appears that the mixing is stronger, leading 
to similar intensities for the two bands.  The pertinent 
value to consider is thus the sum of the $B$6 and 3$d\sigma$0 
oscillator strengths.  This is about 0.0066, a value more in 
line with the $f$-value for the other isotopologues.  
The second exception, $W$1 in $^{13}$C$^{18}$O, is so 
severe that we are not able to derive an $f$-value.  
This band is strongly overlapped by another band of 
unknown origin, whose profile does not resemble a 
$\pi\sigma$ or $\sigma\sigma$ band.  Since its intensity 
varies with pressure as $W$1 does, the band probably 
arises from a valence state perturber.  Thus it is not 
possible to separate its contribution from that of $W$1.  
Finally, we note that the $f$-values for the $W$ bands are 
relatively strong, with the (2-0) band being the 
strongest.  The results for the $W$ bands are 
consistent with the theoretical predictions of 
Cooper \& Kirby (1988) $-$ i.e., $f$-values of about 
10$^{-2}$ and Franck-Condon factors favoring the 
(2-0) band.

Our oscillator strengths are generally consistent 
with earlier determinations.  However, the previous 
absorption experiments using synchrotron sources 
(Eidelsberg et al. 1991; Stark et al. 1991, 1992, 
1993; Yoshino et al. 1995) tend to have somewhat 
smaller $f$-values than ours and those from the 
interstellar study of Sheffer et al. (2003), even 
though the results agree within their mutual 
uncertainties.  Potential problems with optically 
thick absorption noted by Federman et al. (2001), 
Sheffer et al. (2003), and Eidelsberg et al. (2004a) 
are less of an issue for the current set of 
transitions because they are weaker or have 
intrinsic line widths that are broadened by 
predissociation (see below).  The one instance where optical 
depth effects may have played a role involves the 
results for $E$1 of Eidelsberg et al. (1991) since 
predissociation is a minor channel for this band.

A comparison of predissociation rates apppears in Table 2.  
Whenever possible, results for $e$ and $f$ parity are given 
separately.  The $k_J$ term listed for many of the entries 
indicates how the width depends on $J$ 
[$k_p$ $=$ $k_0$~$+$~$k_J$~$J$($J$$+$1)].  Except for the 
results by Eidelsberg et al. (1991), which were based on 
low-resolution spectra of absorption against continua 
from a synchrotron, the other values given in this Table are 
studies involving laser excitation.  For the most part, there is 
general agreement among all results.  Factors of a few difference 
(at most) exist between our widths and the order of magnitude 
evaluations given by Eidelsberg et al. (1991).  Although the 
number of results for comparison with those from laser studies 
is smaller, the agreement is very good.  The most significant 
differences are found with the widths derived by Drabbels et 
al. (1993), especially in light of the fact that the other 
studies yield a set of consistent results.  The close 
correspondence with the laser studies, when available, gives 
us confidence in our other determinations.  These will 
provide modelers of CO photodissociation with a more complete 
set of predissociation rates.  We also note that since the 
rates have values greater than 10$^{10}$ s$^{-1}$, the 
corresponding predissociation lifetimes are 0.1 ns or less, which 
are significantly less than radiative lifetimes ($\sim$ 1 ns).  
Thus, predissociation represents the primary decay channel 
for the upper states involved in the $B$6, $W$0, $W$1, $W$2, and 
$W$3 transitions studied here.

\section{Concluding Remarks}

We presented results for oscillator strengths and 
predissociation rates for Rydberg transitions in 
$^{12}$C$^{16}$O, $^{13}$C$^{16}$O, and $^{13}$C$^{18}$O.  
In particular, the $E-X$ (1-0), $B-X$ (6-0), 
and $W-X$ ($v^{\prime}$~$=$~0,~1,~2,~3-$v^{\prime\prime}$~$=$~0)
bands were studied via absorption of synchrotron radiation.  
This suite of measurements represents the most complete database 
available at the present time for these transitions.  Our 
oscillator strengths agree essentially with earlier determinations 
and expand upon available results for the rarer isotopologues.  
The same applies to the values for predissociation rates, 
but now our larger set of measurements provides $J$-dependent 
rates for most of the transitions in the present study.

Our measurements are especially useful for chemical models 
that incorporate details of CO photodissociation.  Oscillator 
strengths and predissociation rates are now available to 
examine selective isotope photodissociation with greater 
precision.  We note that the significant line widths 
for the $B-X$ (6-0) and $W-X$ ($v^{\prime}$~$=$~0,~1,~2,~3-0)
bands minimize the importance of these transitions to selective 
self shielding, where the more abundant forms of CO are 
protected from dissociation, because optical depths at line 
center cannot reach large values.

With the tools now available to us, we plan to extend our set 
of measurements to other transitions below 1000 \AA.  We will 
utilize the high-resolution beam line on SOLEIL, the 
next-generation synchrotron in France, which is nearing 
completion with first light expected in 2007.  High resolution is 
necessary to extract predissociation widths for the transitions 
important for modeling CO photodissociation in PDRs.

\acknowledgments
This work was supported by NASA grant NAG5-11440 to the
University of Toledo and by the CNRS-PCMI program.  The authors 
acknowledge the support of the LURE-Super ACO facility and the 
SU5 beam line team.  We thank Chris McKee for insightful 
comments about the value of our measurements.

\clearpage
\begin{deluxetable}{cccccccccc}
\tablecolumns{10}
\tablewidth{0pt}
\tabletypesize{\scriptsize}
\tablecaption{CO Oscillator Strengths $\times$ 10$^3$}
\startdata
\hline \hline\\
Band & $\lambda$ (\AA) & Isotopologue & Present & 
E91\tablenotemark{a} & C93\tablenotemark{b} & 
S91,92,93\tablenotemark{c} & Y95\tablenotemark{d} & 
Z97\tablenotemark{e} & S03\tablenotemark{f} \\ 
 & & & 295 K & 295 K & 295 K & 295 K & 20 K & 295 K & ISM \\ \hline
$E$1 & 1051.71 & 1216 & $3.6\pm0.3$ & $2.5\pm0.2$ & $3.53\pm0.35$ 
& $3.0\pm0.3$ & $\ldots$ & $4.67\pm0.66$ & $3.3\pm1.1$ \\
 & 1052.28 & 1316 & $4.2\pm0.5$ & $2.5\pm0.2$ & $\ldots$ 
& $3.1\pm0.3$ & $\ldots$ & $\ldots$ & $\ldots$ \\
 & 1052.81 & 1318 & $3.5\pm0.1$ & $2.5\pm0.2$ & $\ldots$ 
& $\ldots$ & $\ldots$ & $\ldots$ & $\ldots$ \\
$B$6 & 1002.56 & 1216 & $8.3\pm0.7$ & $7.9\pm0.8$ & $\ldots$ 
& $7.8\pm0.8$ & $\ldots$ & $\ldots$ & $\ldots$ \\
 & 1002.81 & 1316 & $7.1\pm0.4$ & $7.3\pm0.7$ & $\ldots$ 
& $\ldots$ & $\ldots$ & $\ldots$ & $\ldots$ \\
 & 1003.21 & 1318 & $5.0\pm0.3$ & $7.3\pm0.7$ & $\ldots$ 
& $\ldots$ & $\ldots$ & $\ldots$ & $\ldots$ \\
$W$0 & 972.70 & 1216 & $16.6\pm1.6$ & $12.1\pm1.2$ &
$\ldots$ & $12.9\pm1.3$ & $13.6\pm2.0$ & $\ldots$ & $\ldots$ \\
 & 972.99 & 1316 & $15.1\pm0.7$ & $13.2\pm1.3$ & $\ldots$ 
& $\ldots$ & $\ldots$ & $\ldots$ & $\ldots$ \\
 & 973.30 & 1318 & $13.8\pm2.0$ & $13.2\pm1.3$ & $\ldots$ 
& $\ldots$ & $\ldots$ & $\ldots$ & $\ldots$ \\
$W$1 & 956.25 & 1216 & $16.0\pm1.3$ & $13.5\pm1.4$ & $\ldots$ 
& $14.8\pm1.5$ & $14.8\pm1.5$ & $\ldots$ & $15.8\pm2.0$ \\
 & 956.29 & 1316 & $16.1\pm2.8$ & $16.1\pm1.6$ & $\ldots$ 
& $\ldots$ & $\ldots$ & $\ldots$ & $\ldots$ \\
 & 956.33 & 1318 & $\ldots$ & $16.0\pm1.6$ & $\ldots$ 
& $\ldots$ & $\ldots$ & $\ldots$ & $\ldots$ \\
$W$2 & 941.17 & 1216 & $30.8\pm2.4$ & $25.8\pm2.6$ & $\ldots$ 
& $30.0\pm3.0$ & $20.4\pm3.1$ & $\ldots$ & $23\pm5$ \\
 & 941.65 & 1316 & $29.1\pm1.3$ & $27.9\pm2.8$ & $\ldots$ 
& $\ldots$ & $\ldots$ & $\ldots$ & $\ldots$ \\
 & 942.16 & 1318 & $28.5\pm2.0$ & $27.9\pm2.8$ & $\ldots$ 
& $\ldots$ & $\ldots$ & $\ldots$ & $\ldots$ \\
$W$3 & 925.81 & 1216 & $19.7\pm1.4$ & $16.3\pm1.6$ & $\ldots$ 
& $14.9\pm1.5$ & $17.0\pm2.6$ & $\ldots$ & $19.8\pm2.4$ \\
 & 927.24 & 1316 & $18.7\pm1.4$ & $18.7\pm1.9$ & $\ldots$ 
& $\ldots$ & $\ldots$ & $\ldots$ & $\ldots$ \\
 & 928.58 & 1318 & $15.4\pm2.4$ & $18.6\pm1.9$ & $\ldots$ 
& $\ldots$ & $\ldots$ & $\ldots$ & $\ldots$ \\
\enddata
\tablenotetext{a}{Eidelsberg et al. 1991.}
\tablenotetext{b}{Chan et al. 1993.}
\tablenotetext{c}{Stark et al. 1991, 1992, 1993.}
\tablenotetext{d}{Yoshino et al. 1995.}
\tablenotetext{e}{Zhong et al. 1997.}
\tablenotetext{f}{Sheffer et al. 2003.}
\end{deluxetable}

\clearpage

\begin{deluxetable}{cccccccccc}
\rotate
\tablecolumns{10}
\tablewidth{0pt}
\tabletypesize{\scriptsize}
\tablecaption{Predissociation Rates (10$^{11}$ s$^{-1}$)}
\startdata
\hline \hline\\
Band & Isotopologue & Parity & Present &
E91\tablenotemark{a} & L92a\tablenotemark{b} &
L92b\tablenotemark{c} & D93\tablenotemark{d} &
E94\tablenotemark{e} & K95\tablenotemark{f} \\ \hline
$B$6 & 1216 & e,f & 
13.0(1.0)~\tablenotemark{g}~$^,$~\tablenotemark{h} & 3.3 &
$\ldots$ & $\ldots$ & $\ldots$ & $\ldots$ & $\ldots$ \\
 & 1316 & e,f & 2.3(0.1) & 3.3 &
$\ldots$ & $\ldots$ & $\ldots$ & $\ldots$ & $\ldots$ \\
 & 1318 & e,f & 33.8(3.6) & 3.3 &
$\ldots$ & $\ldots$ & $\ldots$ & $\ldots$ & $\ldots$ \\
$W$0 & 1216 & e & 0.12(0.01)$+$0.024(0.012)$x$ & 0.1 
& $\ldots$ & 0.36(0.19) & 0.070(0.004)$+$0.0264(0.0006)$x$ 
& 0.10(0.01)$+$0.022(0.002)$x$ & $\ldots$ \\
 & & f & 0.12(0.01) & $\ldots$ & $\ldots$ & $\ldots$ 
& 0.096(0.008) & $\ldots$ & $\ldots$ \\
 & 1316 & e & 0.10$+$0.020$x$~\tablenotemark{i} & 0.1 & $\ldots$
& $\le$0.3 & $\ldots$ & $\le$0.3$+$0.013(0.002)$x$ & $\ldots$ \\
 & & f & 0.10~\tablenotemark{i} & $\ldots$ &
$\ldots$ & $\ldots$ & $\ldots$ & $\ldots$ & $\ldots$ \\
 & 1318 & e & 0.11(0.01)$+$0.01(0.01)$x$ & 0.1 &
$\ldots$ & $\ldots$ & $\ldots$ & $\ldots$ & $\ldots$ \\
 & & f & $\ldots$ & $\ldots$ & $\ldots$ & $\ldots$ & $\ldots$ & 
$\ldots$ & $\ldots$ \\
$W$1 & 1216 & e,f & 7.2~\tablenotemark{i} & 3.3 &
$\ldots$ & $\ldots$ & $\ldots$ & $\ldots$ & $\ldots$ \\
 & 1316 & e,f & 7.2~\tablenotemark{i} & 3.3 &
$\ldots$ & $\ldots$ & $\ldots$ & $\ldots$ & $\ldots$ \\
$W$2 & 1216 & e & 1.0(0.2)$+$0.034(0.013)$x$ & 1.0 
& 1.15(0.15) & 1.0(0.2) & 3.86(0.21) & 1.21(0.10) & $\le$1.9 \\
 & & f & 1.0(0.2)$+$0.018(0.002)$x$ & $\ldots$ & $\ldots$
& $\ldots$ & $\ldots$ & $\ldots$ & $\ldots$ \\
 & 1316 & e & 0.76(0.21)$+$0.042(0.021)$x$ & 1.0 
& 0.58(0.13) & 0.47(0.19) & $\ldots$ & 0.75(0.15) & $\ldots$ \\
 & & f & 0.76(0.21)$+$0.021(0.010)$x$ & $\ldots$ 
& $\ldots$ & $\ldots$ & $\ldots$ & $\ldots$ & $\ldots$ \\
 & 1318 & e & 0.42(0.06)$+$0.038(0.017)$x$ & 1.0 
& $\ldots$ & $\ldots$ & $\ldots$ & $\ldots$ & $\ldots$ \\
 & & f & 0.42(0.06)$+$0.021(0.010)$x$ & $\ldots$ & $\ldots$ 
& $\ldots$ & $\ldots$ & $\ldots$ & $\ldots$ \\
$W$3 & 1216 & e & 1.6(0.6)$+$0.13(0.02)$x$ & 3.3 
& $\ldots$ & $\ldots$ & $\ldots$ & $\ldots$ & $\ldots$ \\
 & & f & 3.6(0.3)$+$0.040(0.004)$x$ & $\ldots$ & $\ldots$ 
& $\ldots$ & $\ldots$ & $\ldots$ & $\ldots$ \\
 & 1316 & e & 1.0(0.1)$+$0.22(0.07)$x$ & 3.3 
& $\ldots$ & $\ldots$ & $\ldots$ & $\ldots$ & $\ldots$ \\
 & & f & 3.1(0.7)$+$0.072(0.013)$x$ & $\ldots$ &
$\ldots$ & $\ldots$ & $\ldots$ & $\ldots$ & $\ldots$ \\
 & 1318 & e,f & 4.4~\tablenotemark{i} & 3.3 & $\ldots$ 
& $\ldots$ & $\ldots$ & $\ldots$ & $\ldots$ \\
\enddata
\tablenotetext{a}{Eidelsberg et al. 1991.}
\tablenotetext{b}{Levelt et al. 1992a.}
\tablenotetext{c}{Levelt et al. 1992b.}
\tablenotetext{d}{Drabbels et al. 1993.}
\tablenotetext{e}{Eikema et al. 1994.}
\tablenotetext{f}{Komatsu et al. 1995.}
\tablenotetext{g}{When the predissociation rate is known to
have a $J$ dependence, the value is given by $k_p$ $=$
$k_0$~$+$~$k_Jx$, where $x$ is $J$($J+1$).}
\tablenotetext{h}{Uncertainties given in parentheses.}
\tablenotetext{i}{Held fixed during the profile synthesis.}
\end{deluxetable}

\clearpage

\begin{figure}
\begin{center}
\includegraphics[scale=1.5]{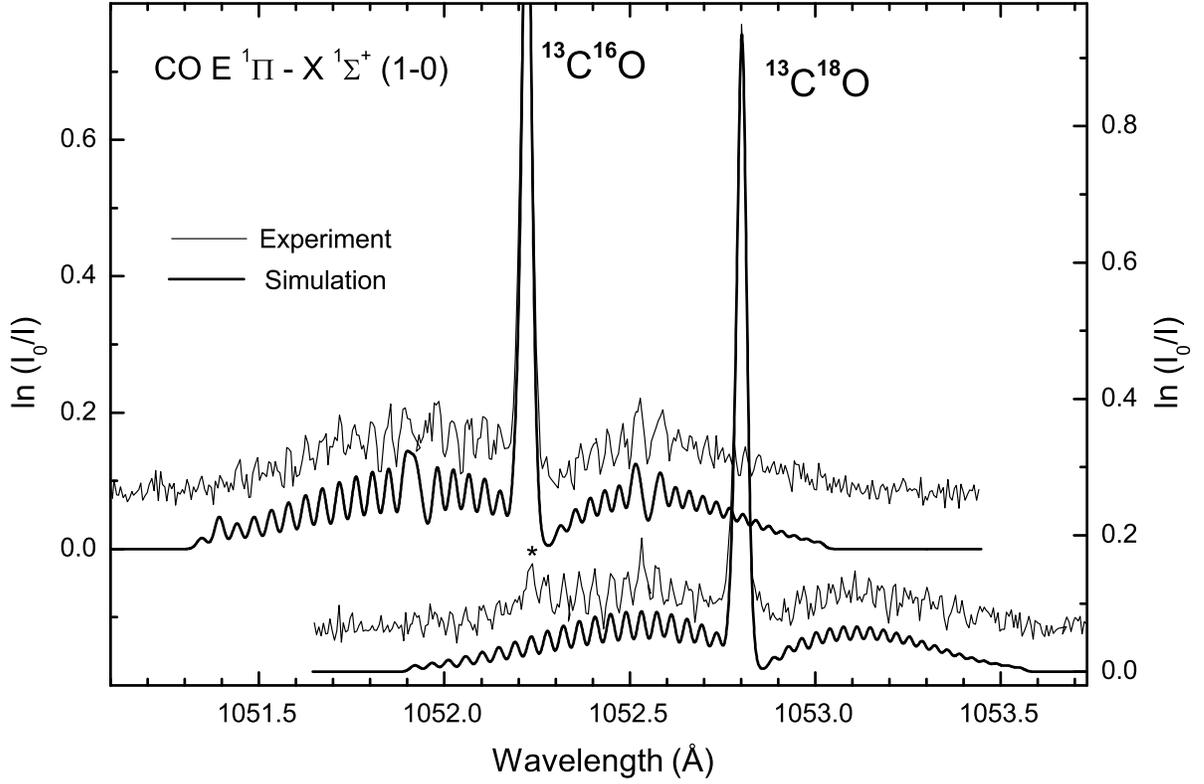}
\end{center}
\vspace{0.3in}
\caption{Experimental and best fitting simulated spectra of 
the $E$1 band for $^{13}$C$^{16}$O and $^{13}$C$^{18}$O.  
The experimental spectra have been shifted vertically for 
clarity here and in Figures 2 and 3.  The scale on the right 
hand side refers to the spectrum of $^{13}$C$^{18}$O.  
Note the local perturbations at $J$ $=$ 7 for 
$^{13}$C$^{16}$O.  The feature marked by an asterisk is due to 
the impurity $^{13}$C$^{16}$O.}
\end{figure}

\clearpage

\begin{figure}
\begin{center}
\includegraphics[scale=1.5]{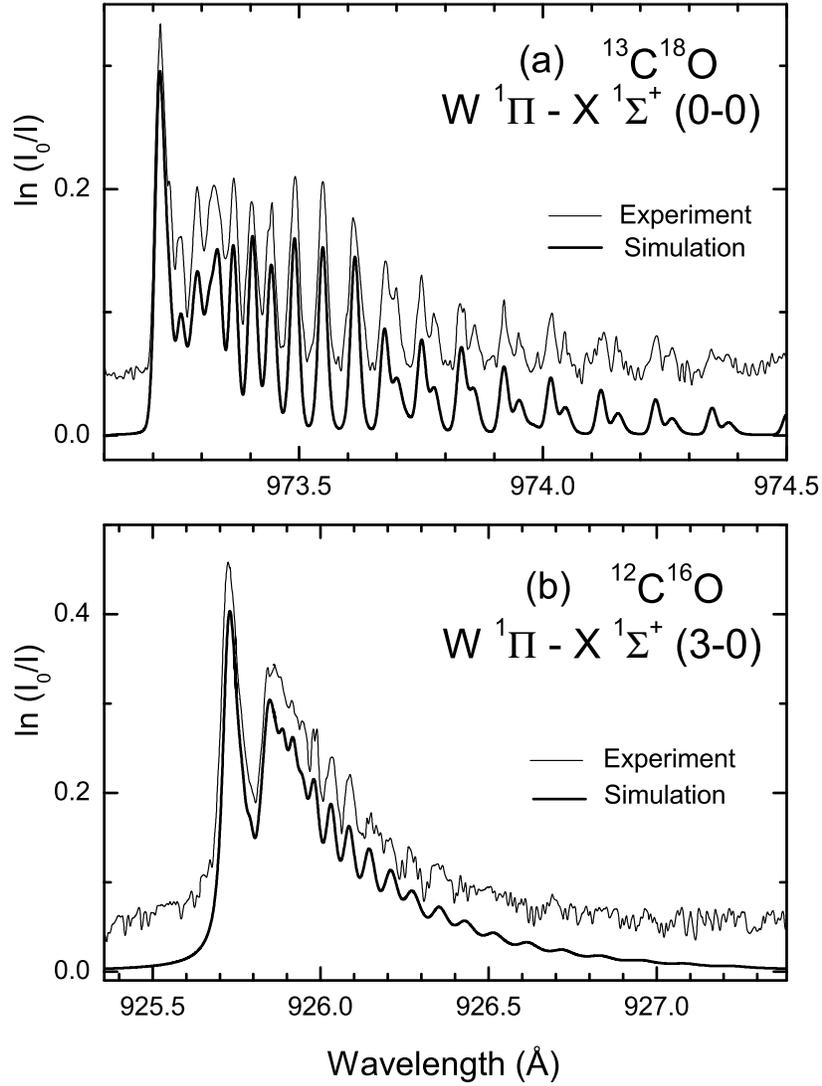}
\end{center}
\vspace{0.3in}
\caption{Experimental and best fitting simulated spectra of the 
$W$0 band for $^{13}$C$^{18}$O (a) and the $W$3 band for 
$^{12}$C$^{16}$O (b).  The considerably broader lines in $W$3 
reveal stronger predissociation (see data in Table 2).}
\end{figure}

\clearpage

\begin{figure}
\begin{center}
\includegraphics[scale=1.5]{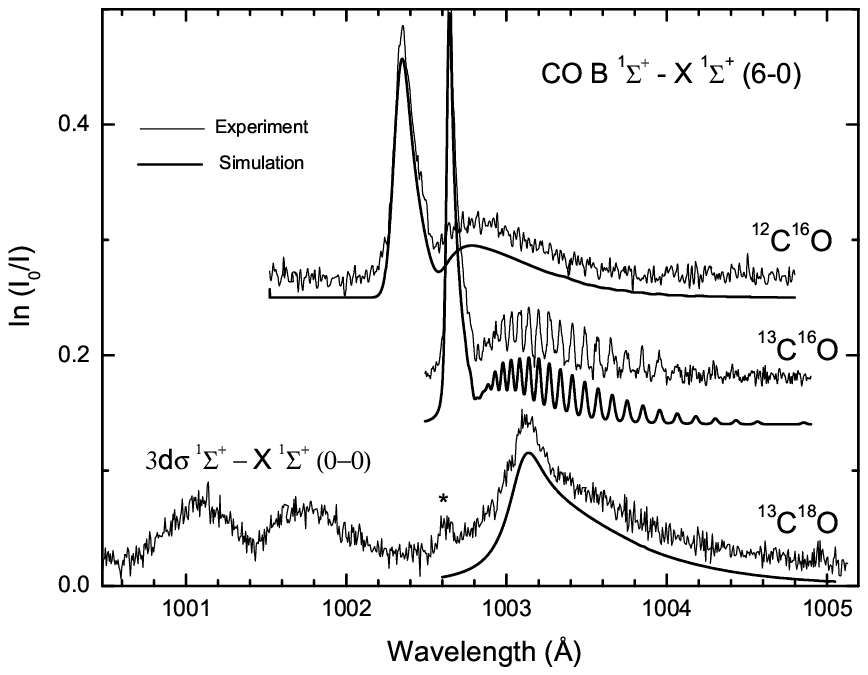}
\end{center}
\vspace{0.3in}
\caption{Experimental and best fitting simulated spectra of the 
$B$6 band for $^{12}$C$^{16}$O, $^{13}$C$^{16}$O and 
$^{13}$C$^{18}$O.  The spectra for $^{12}$C$^{16}$O and 
$^{13}$C$^{16}$O have been shifted upward for clarity.  
The feature marked by an asterisk is due to 
the impurity $^{13}$C$^{16}$O.  The 3$d\sigma$0 forbidden band also 
appears in the $^{13}$C$^{18}$O spectrum.  For this isotopologue, 
the upper state is strongly mixed with $B$6 and borrows about half 
its intensity.}
\end{figure}

\end{document}